\documentstyle[pre,aps,epsf,12pt]{revtex}

\newcommand{\beq}{\begin{equation}}
\newcommand{\eeq}{\end{equation}}
\newcommand{\bea}{\begin{eqnarray}}
\newcommand{\eea}{\end{eqnarray}}

\begin{document}
\pagestyle{empty}
\baselineskip=24pt

\title{ Neutron Stars and Quantum Billiards }

\author{ Aurel BULGAC $^{1,2}$ and Piotr MAGIERSKI $^{1,3}$}  

\address{ $^1$Department of Physics,  University of Washington,
          Seattle, WA 98195--1560, USA }
\address{ $^2$ Max--Planck--Insitut f\"ur Kernphysik, Postfach 10 39
          80, 69029 Heidelberg, GERMANY}
\address{ $^3$ Institute of Physics, Warsaw University of Technology,
          ul. Koszykowa 75, PL--00662, Warsaw, POLAND } 
\address{ e--mail:  bulgac@phys.washington.edu and magiersk@if.pw.edu.pl }


\maketitle

\begin{abstract}

Homogeneous neutron matter at subnuclear densities becomes unstable
towards the formation of inhomogeneities. Depending on the average
value of the neutron density one can observe the appearance of either
bubbles, rods, tubes or plates embeded in a neutron gas. We estimate
the quantum corrections to the ground state energy (which could be
termed either shell correction or Casimir energy) of such phases of
neutron matter.  The calculations are performed by evaluating the
contribution of the shortest periodic orbits in the Gutzwiller trace
formula for the density of states. The magnitude of the quantum
corrections to the ground state energy of neutron matter are of the
same order as the energy differences between various phases.

\end{abstract}


\vspace{0.5cm}

It is somewhat surprising to find out that the problem we are going to
present here, which could have naturally emerged as a typical study
case in quantum chaos, has somehow eluded the attention of this
community. Various billiard problems, in particular the Sinai billiard
in 2D, are paradigmatic in quantum chaos studies. The structures
believed to appear in the crust of neutron stars, namely bubbles,
tubes, rods and plates, known also as the "pasta phase" \cite{baym},
are nothing else but one of several natural realizations of Sinai
billiards. The energetics of these and similar phases at higher
densities involving quarks has been studied for many years \cite{baym}
(this reference list is not meant to be exhaustive).

All the studies mentioned above have estimated the ground state energy
of neutron matter only in the liquid drop or Thomas--Fermi
approximation. In this approximation applied to finite fermi systems
-- such as an atom, a nucleus, a metallic cluster or a quantum dot --
one fails to put in evidence the fact that in the case of closed
shells these systems show an enhanced stability. Apparently, an
agreement has been reached in literature concerning the existence of
the following chain of phase changes as the density is increasing:
nuclei $\rightarrow$ rods $\rightarrow$ plates $\rightarrow$ tubes
$\rightarrow$ bubbles $\rightarrow$ uniform matter. There are a couple
of studies, where the role of quantum corrections has been taken into
account, one being the Hartree--Fock calculations of
Refs. \cite{negele}. In Ref. \cite{oya2} the shell effects due to the
bound nucleons in nuclei immersed in a neutron gas only (mainly
protons) have been computed.  There it was concluded that this type of
quantum corrections to the ground state energy will not lead to any
qualitative changes in the sequence of the nuclear shape transitions
in the neutron star crust. However, no other type of quantum
corrections was suggested.  We have shown that the quantum corrections
to the ground state energy of neutron matter arising from the
unbounded motion of fermions outside the bubbles in particular, are
more important quantitatively and thus can lead to qualitative changes
in the physics of neutron matter at subnuclear densities
\cite{piotr}. In particular, since the magnitude of the quantum
corrections to the ground state energy is of the same order of
magnitude as the energy differences between various ``pasta phases'',
there is no reason to expect that the sequence ``nuclei $\rightarrow$
rods $\rightarrow$ plates $\rightarrow$ tubes $\rightarrow$ bubbles
$\rightarrow$ uniform matter'' is occurring and instead strong
disorder could dominate over various regular lattices.

Let us start our analysis with reviewing a somewhat simpler case, that
of systems with only one bubble \cite{bub1,bub2}. About half a century
ago it was predicted that if a large nucleus was ever to be observed,
most likely it would have either the shape of a torus or of a sphere
with an empty spherical cavity inside (this is what is typically
referred to as a bubble nucleus) \cite{wheeler}. Such geometries lead
to a minimum of the sum of Coulomb and surface energies. Recently it
was suggested that many fermion systems with a similar geometry can be
created in the case of highly charged atomic clusters
\cite{dietrich}. In order to focus our attention on the quantum
corrections, we shall discard here the role of long range Coulomb
interaction, which is easy to account for, and we shall sidestep the
issue of bubble stability as well and refer the interested reader to
earlier literature. In the case of atomic clusters the Coulomb energy
plays a secondary role anyway \cite{bub2}.  For the sake of
simplicity, we shall consider only one type of fermions with no
electric charge.

It is known that in the case of a saturating $N$--fermion system the
total energy has the general structure
\begin{equation} E(N)=e_vN +e_sN^{2/3}+e_cN^{1/3} + E_{sc}(N).
\label{eq:liq} \end{equation}
The first three terms represent the smooth, liquid drop part of the
total energy. $E_{sc}(N)$ is the pure quantum contribution, known as
shell correction energy, the amplitude of which grows in magnitude
approximately as $\propto N^{1/6}$, see Ref. \cite{strutin}. The
liquid drop part of the energy in the case of billiard systems is referred
to as the Weyl energy \cite{weyl,brack}.

One of the simplest questions one can raise is: How should one determine
the position of a bubble or void? Considerations based on naive
symmetry arguments have been used in the past, without any critical
consideration, to simply position the bubble in the very
center of the system. Recently we have shown that moving the bubble
off--center can often lead to a greater stability of the system, due to
shell correction energy effects alone \cite{bub1,bub2}. It is worth
noting at this point that the shell correction energy bears a
remarkable similarity to the Casimir energy in quantum field theory
and in critical phenomena \cite{cas,fish,casimir}. The Casimir energy
can be thought of as a specific measure of the magnitude and nature of
the fluctuations induced in the energy spectrum by the presence of
various ``obstacles'' and is computed with a formula very similar to
the one used to estimate the shell correction energy \cite{piotr}.
There is a large number of studies of various fermion systems, which
shows that the character of the shell corrections is strongly
correlated with the existence of regular or chaotic motion in the
classical limit \cite{brack,balian,ben,strut}. Integrable systems have
conspicuous large gaps in their spectra, while chaotic systems have a
spectrum characterized by rather small fluctuations and essentially no
large gaps. If a spherical bubble appears in a spherical system and
if the bubble is positioned at the center, then for certain magic
fermion numbers the shell correction energy $E_{sc}(N)$, and hence the
total energy $E(N)$, has a very deep minimum. However, if the number
of particles is not magic, in order to become more stable the system
will in general tend to deform.  Real deformations lead to an
increased surface area and thus to an increased liquid drop energy. In
the case of systems with bubbles/voids there is a somewhat unexpected
extra soft deformation mode -- bubble positioning.  Merely shifting a bubble
off--center deforms neither the bubble nor the external surface and
therefore, the liquid drop part of the total energy of the system
remains unchanged. Therefore, one would expect that classically this
deformation mode has exactly zero energy. Quantum mechanics at first
glance seems to lend some support to this idea.  In recent years it
was shown that in a 2--dimensional annular billiard, which is the
2--dimensional analog of spherical bubble nuclei, the motion becomes
more chaotic as the bubble is moved further from the center
\cite{bohigas}.  One might anticipate then that the importance of the
shell corrections diminishes when the bubble is off--center and thus
becomes easier to move a bubble, the further from the
center it is. One might also naively assume that the
most favorable arrangement is the one with the bubble at the very
center, when the system has the highest symmetry.  It came as somewhat
of a surprise to find out that this is not the case most of the time
and that the profile of the energy surface has a very unexpected
structure \cite{bub1,bub2}. 

We shall refer the reader to our previous work for more details
concerning the physics of systems like bubble nuclei and revert now to
the case of bubbles, rods, tubes and plates floating in an infinite
neutron gas. All calculations we present here for infinite systems
were performed at constant chemical potential.  In order to better
appreciate the nature of the problem we present here, let us consider
the following situation. Let us imagine that two spherical identical
bubbles have been formed in an otherwise homogeneous neutron
matter. One can then ask the following apparently innocuous question:
``What determines the most energetically favorable arrangement of the
two bubbles?''  According to a liquid drop model approach the energy
of the system should be insensitive to the relative positioning of the
two bubbles. A similar question was raised in condensed matter
studies, concerning the interaction between two impurities in an
electron gas. In the case of two ``weak'' and point--like impurities
the dependence of the energy of the system as a function of the
relative distance between the two impurities ${\bf a}$ is given by
\begin{equation}
E({\bf a})=\frac{1}{2}\int d{\bf r}_{1} \int d{\bf r}_{2}
V_{1}({\bf r}_{1})\chi ({\bf r}_{1}-{\bf r}_{2}-{\bf a})V_{2}({\bf r}_{2}),
\end{equation}
where $\chi ({\bf r}_{1}-{\bf r}_{2}-{\bf a}) $ is the Lindhard
response function of a homogeneous Fermi gas and $V_{1}({\bf r}_{1})$
and $V_{2}({\bf r}_{2})$ are the potentials describing the interaction
between impurities and the surrounding electron gas.  At large
distances, when $k_{F}a \gg 1$, this expression leads to an
interaction first derived by Ruderman and Kittel \cite{rk,fw}:
\begin{equation} \label{r-k}
E ({\bf a}) \propto \frac{\cos (2k_{F} a)}{a^{3}} ,
\end{equation}
where $k_{F}$ is the Fermi wave vector and $m$ is the fermion mass.
This asymptotic behavior is valid only for point--like impurities,
when $k_{F}R \ll 1$, where $R$ stands for the radius of the two
impurities. This condition is typically violated for nuclei and
bubbles immersed in a neutron gas, for which $k_FR\gg 1$. We have
shown in Ref. \cite{piotr} that in the case of large ``impurities''
($k_FR\gg 1$) and at constant chemical potential the interaction
energy at large separations $a$ becomes
\beq \label{bb}
E_{\circ \circ} \approx \frac{\hbar ^2 k_F^2}{2m}
\left ( \frac{R}{a} \right ) ^2 \frac{2\sin (2k_F a)}{\pi ^2} .
\eeq
The most striking aspect of this interaction is the fact that it
is inversely proportional to the square of the separation,
instead of $1/a^3$ as in Eq. (\ref{r-k}). The bubble--bubble
interaction has a surprinsingly long range. Apart from Coulomb and
gravitational interactions, there are no other longer ranged
interactions known between two objects. One can advance a simple
qualitative explanation of the difference in the power law exponent of
the impurity--impurity interaction (\ref{r-k}) and the bubble--bubble
interaction (\ref{bb}) in terms of waves. A large ``obstacle''
reflects back more of the incident wave than a point object, which
acts like a pure $s$--wave scatterer.  We shall see below that
perfectly planar surfaces lead to an interaction which decays with
distance even slower, inversely proportional to the separation between
surfaces.  Naturally, when the incident and reflected waves interfere
constructively, one does expect a more energetically favorable
geometric configuration.  Obviously, the mere fact that the
interaction (\ref{bb}) oscillates shows that for each value of $a=
\pi(n+3/4)/k_F$, where $n$ is an integer, the potential has a minimum
and one thus expects that a bubble--bubble molecule with such a radius
can be formed. Moreover, for each such size various vibrational and
rotational states, corresponding to small fluctuations of the
bubble--bubble distance and angular velocity are most likely to
exist. Tunneling between various molecular sizes should also appear.
However, the treatment of such molecular systems and tunneling should be
deferred until the bubble inertia is also estimated.

The arguments of the cosine in Eq. (\ref{r-k}) and of the sine in
Eq. (\ref{bb}) are nothing else but the classical action in units of
$\hbar$ of the bouncing periodic orbit between the two impurities and
this suggests the most likely method of attack for this problem, the
semiclassical approximation based on the Gutzwiller trace formula for
the density of states. As a matter of fact this is the way
Eq. (\ref{bb}) has been derived \cite{piotr}.  The formation of
various inhomogeneities in an otherwise uniform Fermi gas can be
characterized by several natural dimensionless parameters: $k_Fa\gg
1$, where as above $a$ is a characteristic separation distance between
two such inhomogeneities; $k_FR\gg 1$, where $R$ is a characteristic
size of such an inhomogeneity; and $k_Fs  = {\cal O}(1)$, where $s$ is a
typical ``skin'' thickness of such objects. The fact that the first
two parameters, $k_Fa$ and $k_FR$, are both very large makes the
adoption of the semiclassical approach natural. Since the third
parameter, $k_Fs$, is never too large or too small, one might be
tempted to discard the semiclassical treatment in the case of
saturating fermion systems altogether as unreasonable. However, there
is a large body of evidence pointing towards the fact that even though
this last parameter in real systems is of order unity, the seemingly
unreasonable approximation $k_Fs\ll 1$, which we shall adopt in this
work, is surprisingly accurate \cite{brack}.  The corrections arising
from considering $k_Fs = {\cal O}(1)$ lead mainly to an overall energy
shift, which is largely independent of the separation among various
objects embedded in a Fermi gas.  On one hand, this type of shift can
be accounted for in principle in a suitably implemented liquid drop
model or Thomas--Fermi approximation. On the other hand, the
semiclassical corrections to the ground state energy arising from the
relative arrangement of various inhomogeneities have to be computed
separately, as they have a different physical nature. We are thus lead
to the natural assumption that a simple hard--wall potential model for
various types of inhomogeneities appearing in a neutron fermi gas is a
reasonable starting point to estimate quantum corrections to the
ground state energy, see Refs. \cite{bub1,bub2,brack} and earlier
references therein.  One might expect that such simplifications will
result in an overestimation of the magnitude of quantum corrections to
the ground state energy, but that the qualitative effect would be
reproduced.

One can compute the quantum corrections to the ground state energy
analytically for plate--like nuclei or voids with the neutron gas
filling the space between slabs.  The quantum correction to the energy
for such a system is given by:
\begin{equation}
\frac{E_{shell}}{L^3}=\frac{E - E_{Weyl}}{L^{3}},
\end{equation}
where the exact and the Weyl (smooth) energy \cite{weyl,brack} per unit
volume are given by
\bea
 \frac{E}{L^3} &=&\frac{2}{L^3}\frac{\hbar^{2}}{2 m a^{2}}
\frac{\pi^{3} }{2}\left ( \frac{L}{a}\right ) ^{2}
\Bigg{[}\frac{1}{4} \left ( \frac{k_{F} a}{\pi} \right
 )^{4} N - 
\frac{N(N+1)(2N+1)(3N^{2}+3N-1)}{120} \Bigg{]}, \\
\frac{E_{Weyl}}{L^{3}}&= &\frac{2}{L^3}
\frac{\hbar^{2}}{2 m a^{2}}
\frac{\pi^{3} }{2}\left ( \frac{L}{a}\right ) ^{2}
\Bigg{[}\frac{1}{5} \left ( \frac{k_{F} a}{\pi} \right ) ^{5}
- \frac{1}{8} \left ( \frac{k_{F} a}{\pi} \right ) ^{4}  \Bigg{]},
\end{eqnarray}
and where
\begin{equation}
N={\mathrm {Int}}\left [ \frac{ k_{F} a}{\pi}   \right  ]
\end{equation}
stands for the integer part of the argument in the square brackets,
and $a=L - 2 R$ is the distance between slabs and $R$ is half of
the width of the slab.  Here $L^3$ is the volume of an elementary
(cubic) cell and the factor ``2'' in front stands for the two spin
states.  Using these formulas, one can show that the shell correction
energy behaves as
\beq
\frac{E_{shell}}{L^3}= \frac{\hbar^2k_{F}^{3}}{48 aL m}
G\left  ( \frac{k_Fa}{\pi} \right ) ,
\eeq
where $G(x)$ is an approximately periodic function of its argument.
For $x\ge 1$, $G(x+1)\approx G(x)$, with properties $G(x=n)\approx
-1$ and $-1\le G(x)\le 0.5$. Note that at small separations $a$ the
shell correction energy is attractive in character.

In the case of rods or tubes and bubbles we resorted to the Gutzwiller
trace formula to estimate the corrections to the density of states. We
were interested in what in the nuclear physics lingo would be called
the ``gross shell structure'' and we discarded the fine details.  As
this approach implies a certain amount of spectral averaging one needs
to account only for the existence of the shortest periodic
orbits. As we are dealing with an infinite system there are no
discrete levels and correspondingly no shells. However, the presence
of ``obstacles'' induces fluctuations in both spatial matter
distribution and density of states, which thus lead to the
appearance of what we term shell correction or Casimir energy.  For
the case of spherical voids there are $26$ periodic orbits between
the nearest neighbors of three different lengths $2L_{1}=2(L-2R)$,
$2L_{2}=2(L\sqrt{2}-R)$ and $2L_{3}=2(L\sqrt{3}-R)$ in the case of a
simple cubic lattice. Thus the shell correction/Casimir energy and
density are equal to:
\begin{eqnarray}
\frac{E_{shell}}{L^3}&=&\frac{1}{L^{3}}\int_{0}^{\mu}\varepsilon
\sum _{i=1}^3  A_i g_{shell}(\varepsilon ,L_i)
d\varepsilon , \\
\rho_{out}&=&\rho_{Weyl} + \frac{1}{L^3}\int_{0}^{\mu}
\sum _{i=1}^3 A_i g_{shell}(\varepsilon ,L_i)
d\varepsilon .
\end{eqnarray}
The contribution due to one periodic orbit to the fluctuating part of
the level density reads:
\begin{equation}
g_{shell}(\varepsilon ,L_i )=
\frac{m L_{i}}{2\pi\hbar^2 k}\sum_{n=1}^{\infty}
\frac{\cos(2 n k L_{i})}{\sinh^{2}(n\kappa_{i})},
\end{equation}
where
\bea
\varepsilon &=& \frac{\hbar^2k^2}{2m}, \\
\kappa_i &=&
\ln \left [ 1 +\frac{L_i }{R} +
\sqrt{ \frac{L_i }{R}\left ( \frac{L_i }{R}+2 \right ) } \right ] .
\eea
and hence we get:
\begin{eqnarray}
& &\frac{E_{shell}}{L^{3}}=
\sum_{i=1}^{3}
\frac{\hbar^2A_i}{16m\pi^{2} L_{i}^2L^3} \sum_{n=1}^{\infty}
\frac{ [ 2 n k_{F} L_{i}
 \cos (2 n k_{F} L_{i}) + (2 n^{2} k_{F}^{2} L_{i}^{2} -1)
   \sin (2 n k_{F} L_{i}) ]}
{n^{3} \sinh ^{2}(n\kappa_{i} )},  \\
& &\rho_{out}=\rho_{Weyl}+
\frac{1}{L^3}\frac{1}{4\pi}
\sum_{i=1}^{3} A_{i}
\sum_{n=1}^{\infty}\frac{\sin(2 n k_{F} L_{i} )}
{n \sinh ^{2}(n\kappa_{i} )},
\end{eqnarray}
where $A_{1}=6, A_{2}=12, A_{3}=8$ respectively. Above, the summation
over $n$ is over repetitions of a given orbit and such summation becomes
superfluous when one performs a spectral averaging in order to extract
the ``gross shell structure''. This shell energy can be regarded as the
interaction energy between bubbles immersed in neutron gas.  The
bubble--bubble interaction (\ref{bb}) was obtained from the above
formula for the case of two bubbles only in the asymptotic limit. One
can derive similar formulas for the case of rods \cite{piotr}. The
fact that the interaction energy between various homogeneities is
described in terms of periodic orbits, points to another nontrivial
aspect: the interaction between three or more such objects cannot be
reduced to simple pairwise interactions. Orbits which bounce among
three or more objects would lead to three--body, four--body and so
forth genuine interactions. Such orbits were not yet included in our
analysis, even though they can be accounted for in a straight
forward manner. Since periodic orbits of this kind are typically
longer, one expects their contribution to be masked when one performs
a spectral averaging in order to extract the ``gross shell
structure''.

Lack of space prevents us from presenting more detailed results and
discussing many other aspects of the energetics of inhomogeneous fermi
systems. Hopefully, the few results we have highlighted here will
convince the reader of the richness of these systems and of the
spectacular role played by geometry and the chaotic versus integrable
character of the single--particle dynamics. Besides static properties,
one should expect that the dynamics of such systems will be extremely
rich and challenging to describe.


\end{document}